# Firm-Dispatchable Power and its Requirement in a Power System based on Variable Generation

Stephen R. Clark *, Craig McGregor

Department of Mechanical and Mechatronic Engineering, Stellenbosch University, Stellenbosch, South Africa *(corresponding author: Sclark@sun.ac.za)

**Abstract**

Many countries have commenced a transition from fossil fuel-based electricity generation systems to sustainable systems based on wind and solar generation. It is often noted that the least cost approach would involve a massive scale-up in the building of variable renewables, supported by battery storage and gas peaking plants. The required backup should be firm-dispatchable generation rather than peaking power. The wind and solar generation aspects for this system are clearly defined and understood, however, the term firm-dispatchable power is not defined and the specific requirements are poorly understood. This study seeks to define firm-dispatchable power in this context and its requirement in the sustainable generation system. The study compares 100% renewable generation scenarios from South Africa, Texas, and the UK to demonstrate the requirement for this firm-dispatchable generation. The results indicate that firm-dispatchable generation must be available to replace the renewable generation completely. The required installed capacity for this firm-dispatchable generation does not vary with the distinct demand profiles of the different locations or their comparative renewable generation profiles. It also does not change significantly with the use of energy storage. The usage for this firm-dispatchable generation will vary due to the comparative economics of its use, but the requirement for its installation does not change.

**Introduction**

South Africa, as most countries in the world, has commenced a transition from a fossil fuelled electricity generation system to one based on solar and wind generation. The intent of this transition is to meet international greenhouse gas emission reduction targets as well as a recognition that these systems are lower cost than conventional fossil fuelled systems. As noted in an editorial in South Africa's Creamers Engineering News on June 30 2023, "*the Presidential Climate Commission (PCC) released its recommendations on South Africa's electricity system, which indicated that a least cost approach would involve a massive scale-up in the building of variable renewables, supported primarily by battery storage and gas peakers*" (Creamer, 2023). This is a concept that has been considered by the international community to be true in most locations (Jacobson et al., 2015; Roy, Sinha & Shah, 2020; Jain, 2023). While this conclusion can be supported by our analysis, the required backup generation should be "firm-dispatchable power" rather than "peakers."

Several studies have demonstrated that an electricity grid based on wind and solar generation with backup firm-dispatchable power is the lowest cost option for a sustainable generation system (Wright et al., 2017; Renné, 2022). While the wind and solar generation aspects for this system are clearly defined and understood, the term firm-dispatchable power is not defined and the specific requirements are poorly understood. Each of the terms, firm generation and dispatchable generation are often discussed, but how these fit together into firm-dispatchable power to support the wind and solar based system has had minimal discussion. This paper seeks to define firm-dispatchable power in this context and its requirement in the sustainable generation system. This definition is supported by technical analysis of three regional grid systems.

**Background**

A traditional fossil fuel-based electricity generation system is based on a combination of baseload, mid-merit, and peaking power plants, as indicated in Figure 1 (Enerdynamics, n.d.). The baseload plants are designed to provide most of the generation running at a constant output. The daily variability in the demand profile is met by adding generation from mid-merit plants and peaking plants if required. In a traditional power system, each of these generation sources is a firm supply source, which can be utilized as much as is needed to balance variable demand.

Most thermal power plants, whether fossil fuelled, nuclear fuelled or renewable systems such as geothermal power and biofuel plants are power-cost-dominated systems because of their high relative capital cost. Once a power plant of a given size is installed, it can provide power up to this design capacity for as many hours as needed. The only significant marginal cost is that of fuel. According to the US EIA, for a coal fuelled power plant, approximately 28% of the cost of produced energy, defined by the levelized cost of electricity (LCOE), is due to fuel cost. For nuclear, the fuel cost is only about 12% of the LCOE (US EIA, 2023a). Because of the high percent of cost being the capital costs, thermal plants are used for as many hours as possible to reduce the per unit cost of generation.

Thermal plants are slower to ramp up and ramp down than typical peaking plants (Kumar et al., 2012). These limits plus the relative high capital cost versus operating cost shift the use of these facilities to use as baseload power. Because of this utilization as firm baseload generation, their development and use competes with rather than complements variable renewable sources (Yuan et al., 2020). Due to economy of scale, thermal plants are generally built at Giga-Watt size or larger and take from seven to twenty years for permitting and construction (Carajilescov & Moreira, 2011; US EIA, 2023a). Thus, they add rigidity to the planning and generation of the overall system. With these limitations, utility scale thermal plants do not have a major position in the low-cost system based on wind and solar generation (Bischof-Niemz, 2019). A recent study from the Royal Society in the United Kingdom (UK) concluded that adding nuclear generation to the supply from wind, solar, storage and firm-dispatchable power increases overall cost of the system (The Royal Society, 2023).

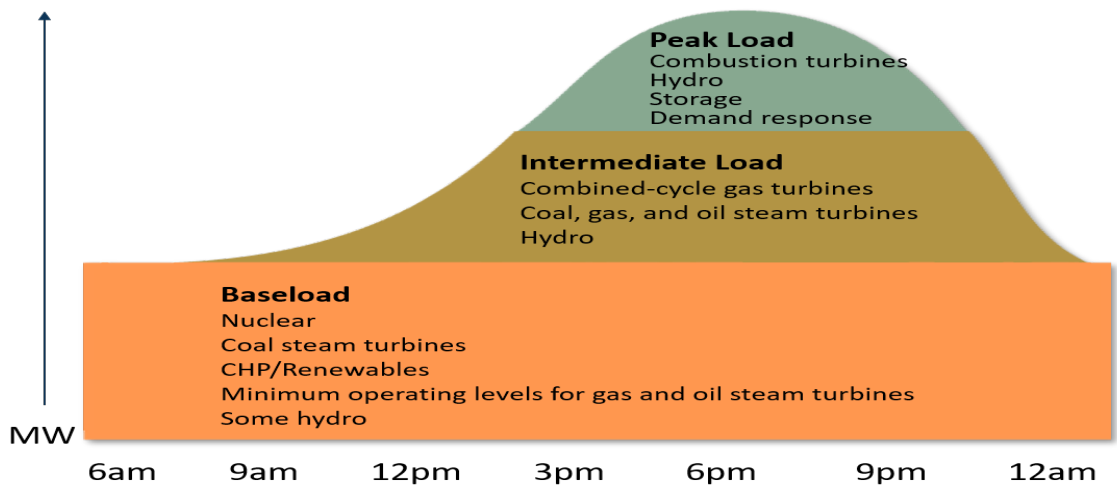

*Figure 1 - Conventual Generation Matrix*

On the other hand, the variability of wind and solar generation are often discussed and are a major concern for generation systems based on these sources (Jacobson et al., 2015; Knorr et al., 2015). The increased need for peaking power is often discussed (Roy, Sinha & Shah, 2020; Jain, 2023). The noted "duck curve" and the related need for power to ramp up quickly and meet the demands of the daily evening peak hours has been reviewed by many sources (Shah & Ahmad, 2019). The effect of this Duck Curve are shown in Figure 2, where the growing requirement for rapid ramp up peaking power with increased use of solar generation is clearly demonstrated. This use of peaking power can supply the power for daily balancing of the system, where power can be added quickly into the system for a few hours in a day to meet the evening peak demand. The need to balance the duck curve was one of the first discussions of peaking power to balance the variable supply solar generation.

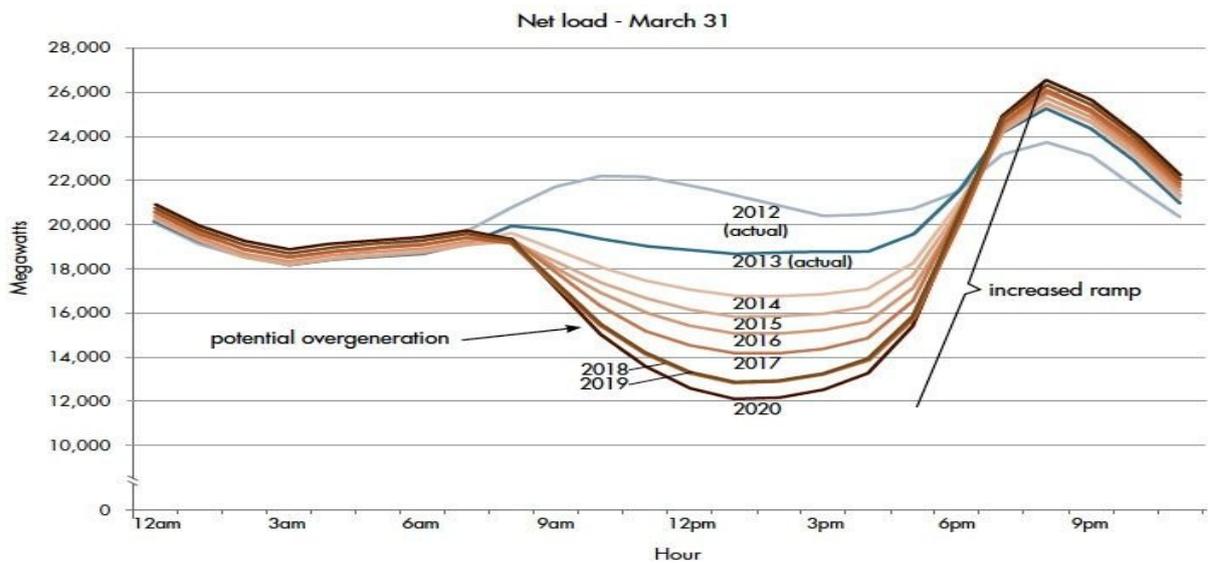

*Figure 2 - California Duck Curve*

The desire to meet this need for daily balancing from renewable sources has resulted in numerous suggestions of meeting this requirement using energy storage mechanisms. These include the use of thermal storage in concentrated solar power (CSP) systems as well as wind and solar generation systems with battery electric storage (Teske et al., 2016; Spector, 2019). Pumped hydro, compressed air storage and various other energy storage are also discussed as potential options to meet this requirement. Power ramp up time is one of the major factors defining the usefulness of these systems for peaking power usage. The major limitation for these systems is that the cost for these systems is generally an energy-based cost rather than a power-based cost. For example, in battery systems, doubling the output time, the energy cost, effectively doubles the overall cost (Cole & Frazier, 2019). This energy versus power cost relationship is true for most storage options.

For nominal peaking power daily usage, this limitation does not significantly add to the cost of these systems. These systems would only be used a few hours in a day and the ramp up and ramp down speed is more significant as well as the ability to meet frequent cycle usage. However, use on a multi-day basis becomes quite expensive as the system must add additional units for each added period of usage (Cole & Frazier, 2019). From previous analysis, it was found that in South Africa the dispatchable generation requirement can last for some days in certain times of the year in a typical year, as shown in Figure 3 (Clark et al., 2022). During this period, the dispatchable power would effectively provide nearly all the needed generation for most of four to five days, as seen in the red curve in Figure 3. Without this generation, the grid would collapse. This is not possible to deal with using peaking generation, storage or demand management.

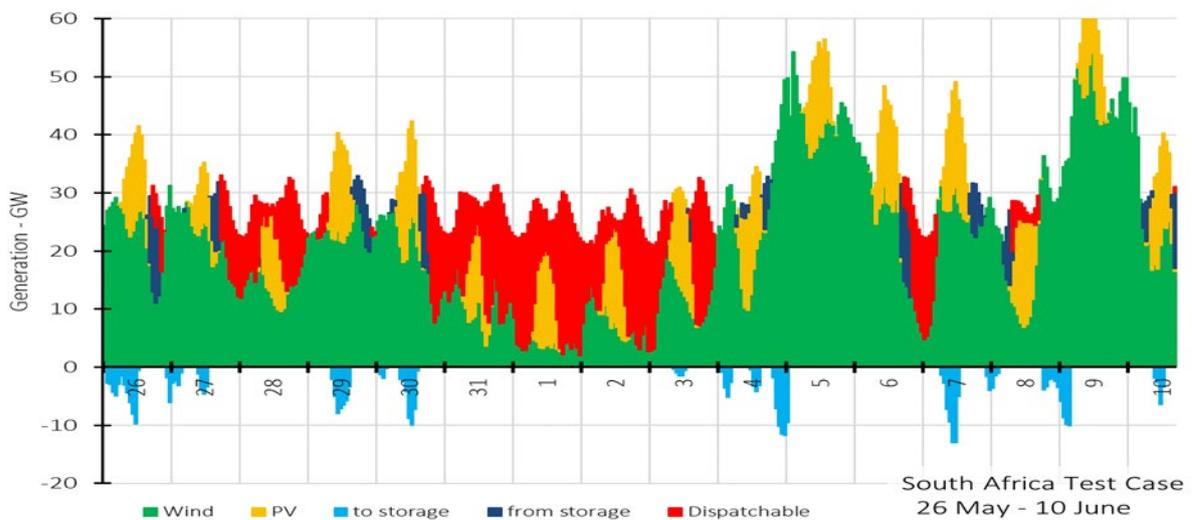

*Figure 3 - South Africa Firm-dispatchable Generation for Hypothetical Case* (Clark et al., 2022)

Reports from Germany, the UK, the USA and Australia have indicated similar time periods of multiple days where firm-dispatchable power is required to provide most of the generation in typical years. In the worst-case scenarios, dispatchable generation could be needed for several weeks to handle "droughts" in wind and solar generation (Brower, 2016; Baraniuk, 2018; Runyon, 2018; Wert et al., 2023). A report from the Royal Society in the UK on energy storage reviewed the need for firm-dispatchable back up power based on 38 weather years in the UK. This report concluded that it would take 1000 times the current pumped hydro storage to provide the needed electricity storage or "far more than could be conceivably supplied by conventional batteries" to balance the UK grid in the worst case scenario (The Royal Society, 2023). IRENA and the IEA have extensively reviewed the time periods over which various storage mechanisms are practical and has shown that options are limited for long term seasonal storage (IRENA, 2017; IEA, 2023)

With peaking generation, energy storage and baseload generation not meeting the specific needs for balancing the solar / wind system for these occasions, another type of backup power is required. This need is for "firm-dispatchable power." This is a term that has not been discussed to any extent in the literature, but due to its requirement to balance the sustainable generation system, its definition in this context and its use must be clarified. This paper attempts to address these two parameters.

**Definition of Firm-dispatchable Generation**

As noted above, the terms "peaking", "dispatchable power" and "firm power" are often discussed in energy literature. The accepted definitions for these terms can be summarised as follows;

**Peaking power:** Enel defines peaking power as - *"A peaking power plant (or "peaker plant" for short) is a power plant that grid operators call on at times of particularly high electricity demand on the grid."*(Enel, 2023).

**Firm power**: The US EIA defines firm power as - "[Firm power is] P*ower or power-producing capacity, intended to be available at all times during the period covered by a guaranteed commitment to deliver, even under adverse conditions.*" (US EIA, 2023b)

**Dispatchable Power**: The University of Calgary defines dispatchable power as - "*A dispatchable source of electricity refers to an electrical power system, such as a power plant, that can be turned on or off; in other words they can adjust their power output supplied to the electrical grid on demand.*" (Univeristy of Calgary, 2023).

While there does not appear to be any discussion of the concept of firm-dispatchable power in the literature, combining the definitions above would give the following definition for the term in the context of a system based on wind and solar generation.

**Firm-dispatchable Power is generating capacity (<u>to replace the wind and solar sources completely</u>) that <u>is always available,</u> that can be turned on or off, or <u>can adjust its power output</u> according to market need.**

**Requirement for Firm-dispatchable Generation**

The premise that is being tested in this analysis is that balancing the power system based on variable sources such as wind and solar requires the installation and usage of firm-dispatchable power. It is expected that while energy storage will have a place in the overall lowest cost mix of generation sources, it will not meet all the backup requirement. This will be demonstrated with the use of scenario modelling for South Africa and comparing the results to that from international analogues.

The requirement for firm-dispatchable generation can be demonstrated in the following hypothetical scenarios. These cases were modelled with the hourly South African grid model as described in the analysis from (Clark et al., 2022). To avoid building in impact for load shedding effects on the demand profile, the South Africa demand profile for 2019 was maintained for this analysis. For the comparison cases from Texas and the UK, 2022 was used as the demand year. The analysis was not intended to compare technology options, but to find the required firm-dispatchable energy requirement. Other potential technology mixes were not considered other than wind, solar and backup power. Costs for all the technologies utilised (wind, solar, storage and dispatchable generation) were based on US NREL cost forecasts for the year 2040. Capital costs assumed were; for wind, 1200 USD per installed kW, for solar PV, 1000 USD/ kW, for dispatchable generation, 800 USD/ kW and for battery storage 200 USD/kWh (US NREL, 2023). Hourly wind and solar generation were based on the capacity factors for the currently installed systems factored up to the assumed level of installed capacity as per the referenced analysis (Clark et al., 2022). In this analysis, it was assumed that all generation is newly installed. For the sake of this comparison, it was assumed that all existing base load generation (coal, nuclear and hydro) have been decommissioned. The only generation sources are wind, solar PV and firm-dispatchable generation. Energy storage based on excess renewable generation was also installed, as economics dictated, for each scenario. Firm-dispatchable power was calculated within the model to balance the supply and demand. These factors were then put through an iteration process to determine the mix of wind, solar PV, battery storage and dispatchable generation to achieve the lowest cost of supply.

Fuel costs for the firm-dispatchable power is one of the major factors in the determination of how much of this energy will be required in the lowest cost scenario. As will be demonstrated in a following section, the impact of fuel cost is not significant on installed capacity requirement, but it has a significant impact on the amount of energy that must be provided to balance the system. A base case 20 USD/GJ fuel cost was assumed for each of the scenarios. Fuel type was not distinguished in this analysis.

**South Africa Test Case**

As shown in the first column of Table 1, the South Africa test case shows that with the above assumptions the system is balanced at the lowest cost with installed wind at 188% of peak demand and solar PV at 70%. The analysis indicates that dispatchable generation must be installed to meet at least 109% of the average demand and 84% of the peak demand.

As indicated in Figure 4 and discussed in detail in the referenced previous analysis, dispatchable generation in South Africa is used throughout the year, but the highest demand for firm-dispatchable generation is in the late autumn to early winter periods – April through June (Clark et al., 2022). This effect was noted with data from each year from 2015 through 2019. This time of the year is the period with long periods of low wind levels. Overall demand for firm-dispatchable energy In the South African system would be in the range of 8% of the overall energy generation in the lowest cost scenario.

In the context of a generation system based primarily on variable sources, the ability to meet demand at all times must be guaranteed through the use of firm-dispatchable generation. The parameters that define this generation source are as follows. These requirements are demonstrated in the South African system studied in this analysis and confirmed in the two international comparisons.

1. Quick ramp up and ramp down. Like the concept of peaking power, firm-dispatchable generation must be able to quickly respond to changes in demand vs supply. As can be seen from the cases studied, the generation from the firm-dispatchable sources must build up to complete supply over short periods to keep the grid in balance.

2. Available to be utilised economically for as long (or short) as needed. This power is expected to be used for minimal periods on an annual basis and generally will only be required for a few hours in a day. There will be periods in a typical year where this power is likely to be required at its full rate for several days. In the worst case, it could be used for periods up to several weeks per use in the maximum usage events.

3. Large Installed Capacities. In a variable wind and solar based system, there are periods within the year where the production from wind and solar sources is effectively zero. To keep the supply / demand in balance, there must be enough firm-dispatchable power to meet all of the demand.

Some of the desired parameters for this type of generation are as follows.

1. Minimum capital cost for installed capacity. As noted in the previous points, it is expected that this generation will be used annually for minimal periods. High capital cost systems with associated significant amortization costs should be avoided as there would not be enough hours of generation to offset the capital costs (Enel, 2023; Eskom, 2023; Wartsila, 2023).

2. Dedicated fuel storage. As this generation source must be available in large volumes for periods of days to weeks in the worst-case scenarios, it should have dedicated fuel storage. It cannot depend on fuel sources with shared priority usage. The requirement for this was covered in the analysis of firm-dispatchable generation in (Clark, Van Niekerk & Petrie, 2020). However, it is possible that multiple types of fuel can be utilised in these plants to meet normal and exceptional demands and fuel storage costs can be optimised with this dual fuel usage (Clark, McGregor & Van Niekerk, 2022).

3. Modular sizing and short implementation times. To avoid under or over building the required generation with associated costs, these two parameters allow the firm-dispatchable generation to be optimally constructed to meet the system needs (World Bank, 2023). Experience from South Africa and internationally shows that dispatchable generation plants can be built within two years and at sizes from less than 100 MW to more than 1500 MW as deemed appropriate (Florida Power and Light, n.d.; Sasol, 2013; Eskom, 2023)

While it might appear that this is an extremely limiting set of parameters, there are several generation sources that meet all these needs. These needs can be met with generation based on combustion engine generators, gas turbine generators or fuel cell generation. Both engine and gas turbine generator plants are utilised in South Africa and around the world (Eskom, 2014; Wartsila, 2019; Siemens, 2020). Engine plants can be built up to several hundred MW capacity and gas turbine plants can be built in modules of several hundred MW each to the size desired (Siemens, 2020; Wartsila, 2023). Fuel cell generation plants are currently more expensive, however as the technology develops, it is expected that their cost will approach the cost of plants using engines or turbines. In addition, fuel cells have a higher efficiency than either of these technologies (Mayyas et al., 2019; Papageorgopoulos, 2019; US EIA, 2023a). Each of these generation sources can be fuelled with an extensive range of fuels depending on the delivered fuel cost (Clark, McGregor & Van Niekerk, 2022).

With fossil fuel usage, these plants will have some greenhouse gas emissions. However, with respect to meeting the overall system greenhouse gas emission target, the elimination of these emissions should be a lower priority target due to the potential use of relatively clean fuels such as natural gas or LPG and their minimal usage (Clark, Van Niekerk & Petrie, 2020; Enel, 2023). Eventually, these plants can be fuelled with green hydrogen and its derivatives, but this should be a secondary target compared to the implementation of the major generation sources, such as wind and solar (IRENA, 2018).

To reduce the amount of firm-dispatchable energy that must be provided, it is possible to build more wind and solar generation and add more storage from batteries or other technologies. However, this comes at an increased cost to the overall system. The amount of required installed firm-dispatchable power does not change significantly within the entire range, until it gets to the zero value.

From this and previous studies, it can be concluded that firm-dispatchable generation is an essential element in the balancing of the South African grid (Bischof-Niemz, 2017; Wright et al., 2017; Clark et al., 2022). It is possible to eliminate the need for this balancing generation by massively over installing wind, solar and storage, but it comes at significant cost to the overall system. The results indicate that while the amount of firm-dispatchable power that must be installed is effectively fixed to completely replace the wind and solar, the amount of energy that it provides is an economic decision.

**Comparison Cases**

To understand how the situation in South Africa compares with international experience it is essential to make comparisons with other systems. Many regional generation systems are integrated into large networks that can use regional balancing to match generation with demand which might mitigate the requirement for firm-dispatchable backup generation to some extent. However, there are examples around the world of systems that are effectively isolated and must balance within their network. Two examples fit the criteria for comparison in size and isolation, Texas and the United Kingdom (UK). The Texas grid operated by the Energy Reliability Council of Texas (ERCOT), at 431 TWh per year, compared to 235 TWh for Eskom, is 187% the size of the South African network operated by ESKOM. However, it has similar wind and solar resources that make comparison appropriate (ERCOT, 2023a). The UK National Grid (NG), with a net demand of approximately 230 TWh of electricity per year, is the same size as the ESKOM grid, but the renewable resources are quite different (Elexon, 2023). Much of the UK wind resource is offshore, but the overall capacity factor of the wind resource in the UK is only 31% compared to the wind capacity factor in South Africa of 36%. The South African wind generation is completely onshore wind, where capacity factors are expected to be lower than offshore wind. The solar power in the UK only has an annual 9% capacity factor compared to 26% in South Africa.

The demand profiles for the three systems are completely different, as can be seen in Figures 4 through 6, with the Texas system dominated by high demand during summer air conditioning periods and the UK higher in the winter period due to heating demands. The South African profile does not have nearly as significant seasonality as either the Texas or the UK system. All three systems have a dominant contribution from wind resources and are subject to the variability of the wind resource, both on a day to day and a seasonal basis.

In this analysis, the same model was built for each of the three systems, based on publicly available data for the year 2022 for Texas and the UK and 2019 for South Africa. The Texas data was from ERCOT published data and the UK was from Elexon data (Elexon, 2023; ERCOT, 2023b). The South African and Texas data are published and utilized on an hourly basis and the UK data is published and utilized on a half-hour basis. Renewable generation was based on 2022 performance factored to the assumed installed capacities. No demand profile growth from 2022 was considered for this analysis. Assumptions for costs and other parameters were as noted above for the South Africa case. For this comparison, it was assumed that fuel cost for dispatchable generation for all three systems was 20 USD/ GJ.

As shown in Table 1, with all the differences within the three systems, the resulting balances are quite similar. Wind resources must be built significantly more than the peak demand – 163% in Texas, 188% in South Africa and 220% in the UK. Solar PV makes a lesser contribution, with installation of 70% of peak demand in South Africa, 79% in Texas but only 11% in the UK. Battery energy storage also makes similar contributions in South Africa and Texas but adds almost no value in the UK system.

From the perspective of the requirement for firm-dispatchable generation, all three systems have similar requirements, with a required installed capacity ranging from 109% for South Africa, 110% in the UK to 118% for Texas based on annual average demand. The usage factor for the dispatchable generation for South Africa and Texas was slightly below 9 % and the UK at 19% for the lowest cost scenario. As can be seen in Figures 4 through 6 for the three cases, while firm-dispatchable generation (represented by the red bars) is required throughout the year, there is significant seasonality to the need. The demand profile for each of the systems is effectively represented as the top of the red bars and all the generation above this line is excess to need. In all three cases there are periods where the firm-dispatchable power is required to meet most of the power for some days to weeks at a time. This is where there is a major challenge to energy storage meeting the demand.

*Table 1 - Firm-dispatchable Generation Comparisons*

|  | South Africa-Eskom | Texas - ERCOT | UK - NG | units |
|---|---|---|---|---|
| **Current Parameters based on 2022** | | | | |
| Annual Demand | 231 | 431 | 230 | TWh |
| Peak Rate | 34 | 80 | 43 | GW |
| Average Rate | 26 | 49 | 26 | GW |
| **Lowest Cost Case Parameters with no Baseload Generation** | | | | |
| Installed Wind | 64 | 130 | 95 | GW |
| Wind Energy | 204 | 392 | 256 | TWh |
| Wind CF | 36 | 34 | 31 | % |
| Wind Percent of Peak Capacity | 188 | 163 | 220 | Percent of Peak Gen. Capacity |
| Installed PV | 24 | 63 | 5 | GW |
| PV Energy | 54 | 129 | 4 | TWh |
| PV CF | 26 | 23 | 9 | % |
| PV Percent of Peak Capacity | 70 | 79 | 11 | Percent of Peak Gen. Capacity |
| Battery Capacity | 13 | 14 | 0 | GW |
| Battery Hours | 4 | 4 | 0 | Hours |
| | | | | |
| Installed Dispatch | 29 | 58 | 30 | GW |
| Dispatch Energy | 21 | 45 | 51 | TWh |
| **Dispatchable Generation Parameters for Lowest Cost Case** | | | | |
| Dispatch CF | 8.6 | 8.9 | 19.3 | % |
| Percent of Peak demand | 84 | 73 | 68 | % |
| Percent of Average demand | 109 | 118 | 110 | % |
| **Renewable Generation Use Parameters** | | | | |
| Renewable Gen | 258 | 521 | 259 | TWh |
| Curtailed Renew. | 48 | 126 | 81 | TWh |
| Percent Curtailed | 18 | 24 | 31 | % |

*Table 1 - Firm-dispatchable Generation Comparisons*

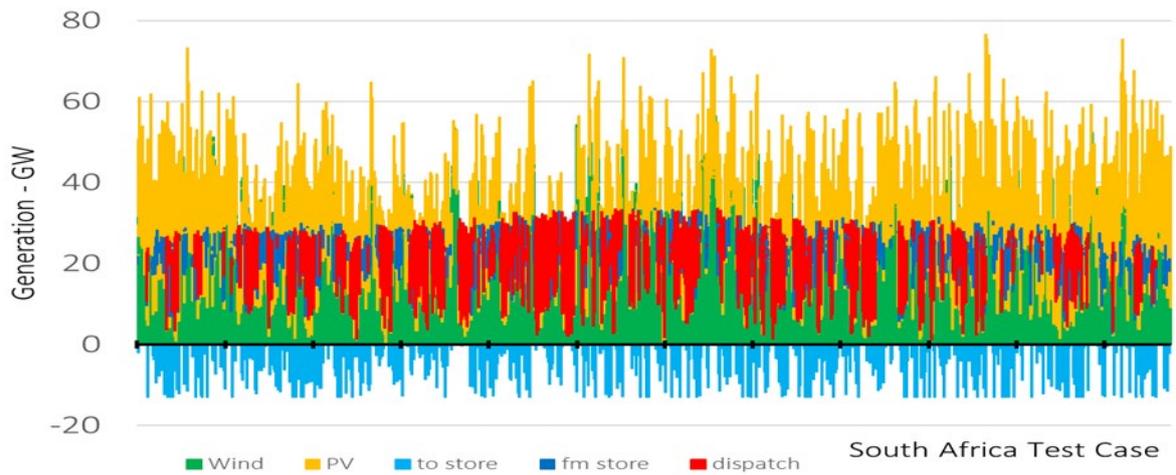

*Figure 4 - South Africa Firm-dispatchable Test Case*

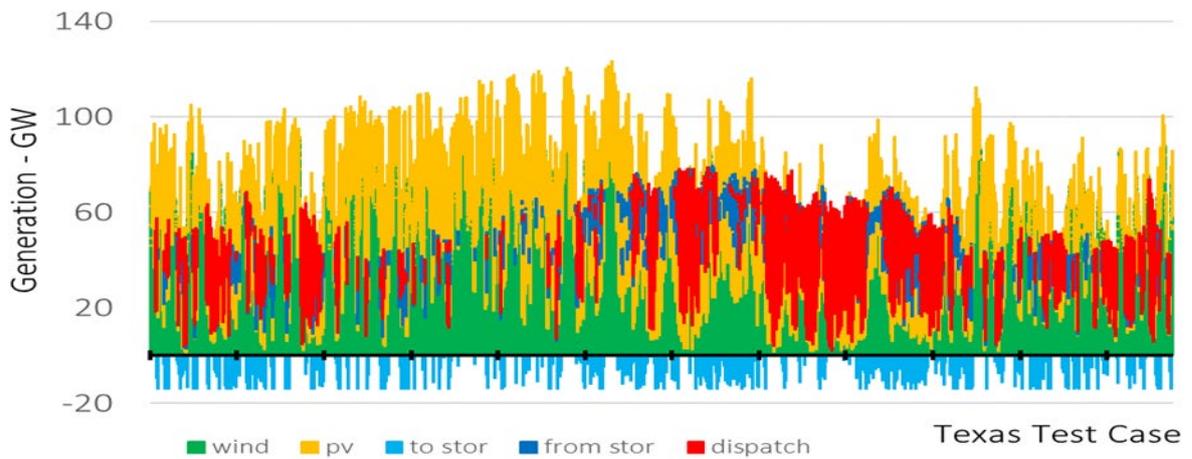

*Figure 5 - Texas Firm-dispatchable Test Case*

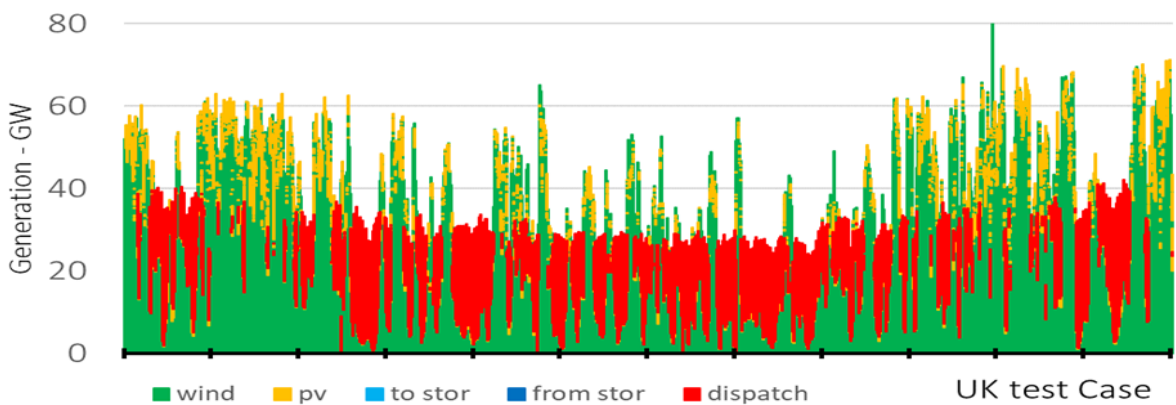

*Figure 6 - UK Firm-dispatchable Test Case*

The three comparison cases all show similar requirements for firm-dispatchable power in proportion to the amount of wind and solar generation. There is the possibility that these numbers might be affected by the cost of storage and on the cost of fuel for the dispatchable generation. In the following sections, these two parameters will be tested.

**Test with low-cost storage**

Energy storage technologies, whether discussing batteries or other proposed alternatives – flow batteries, CSP with thermal storage, compressed air and other proposed concepts is the least developed technology used in this analysis. Because of its level of development, it has the highest level of uncertainty in its potential cost. While the value used of 200 USD/ kWh is less than half of the NREL estimated current cost, it might be considered too conservative (Cole & Frazier, 2019). This leads to the question of the effect of significantly lower storage cost on the amount of firm-dispatchable energy that is required. This could affect both the required installed value and the amount of energy that is required from the installed resource.

Some have argued that just as solar prices dropped so much more than expected, energy storage might do the same and achieve much lower than forecast costs (Goldie-Scot, 2019; Rao, 2021). To test this, it is interesting to determine if extremely low storage costs could eliminate the need for firm-dispatchable power. Test cases for the three systems were run down to storage costs of 10 USD/kWh. In all three scenarios, even with this extremely low storage cost, there was no reduction in the required installed capacity of firm-dispatchable generation. Usage of this firm-dispatchable generation continues to drop as more storage is added, but the peak demand for this firm-dispatchable generation remains. These extreme cases are shown in Figures 8 through 10. Even if the usage period for the firm-dispatchable generation drops to 1%, it would not be possible to operate the grid with a period of days where none of the demand was met. In the worst-case scenarios, this period of non-met demand could be for weeks if firm-dispatchable generation is not available.

**100% Solar PV with Batteries**

Some have argued that solar PV with batteries will become so low cost that the lowest cost generation system will be completely based on solar PV plus batteries (Dorr & Seba, 2020; Mallinson, 2021). This potential case has been analysed in supplement 1. It is technically possible to build enough PV and batteries to completely meet the demand. However, even without economic considerations, the analysis shows that in South Africa the PV required would be 5.7 times that peak demand. For Texas and the UK, the requirement would be higher at 9.1 and 21.4-times peak demand respectively. The battery storage to displace the firm-dispatchable generation would also be quite significant, with the requirement to meet full demand for days with an energy capacity of over 1000 GWh for South Africa to over 3000 GWh for Texas. A significant portion of the wind and solar energy generated would be curtailed unless it could be utilised for other purposes.

The real issue with the solar and battery system would not be so much the cost, but the associated rigidity. The analysis indicates that it would take less than a 1% increase in demand (or decrease in supply) to go from meeting the demand to requiring full firm-dispatchable generation. This could not be accepted for meeting system needs.

*Table 2 - Test Cases with 10 USD / kWh storage costs*

|  | South Africa-Eskom | Texas - ERCOT | UK - NG | units |
|---|---|---|---|---|
| **Current Parameters based on 2022** | | | | |
| Annual Demand | 231 | 431 | 230 | TWh |
| Peak Rate | 34 | 80 | 43 | GW |
| Average Rate | 26 | 49 | 26 | GW |
| **Lowest Cost Case Parameters with no Baseload Generation and very low storage cost** | | | | |
| Installed Wind | 63 | 122 | 102 | GW |
| Wind Energy | 200 | 368 | 274 | TWh |
| Wind Percent of Peak Capacity | 185 | 152 | 238 | Percent of Peak Gen. Capacity |
| Installed PV | 26 | 73 | 4 | GW |
| PV Energy | 59 | 150 | 3 | TWh |
| PV Percent of Peak Capacity | 70 | 91 | 10 | Percent of Peak Gen. Capacity |
| Battery Capacity | 26 | 55 | 67 | GW |
| Battery Hours | 24 | 35 | 21 | Hours |
| Installed Dispatch | 28 | 51 | 30 | GW |
| Dispatch Energy | 7 | 17 | 28 | TWh |
| **Dispatchable Generation Parameters for Lowest Cost Case with very low storage cost** | | | | |
| Dispatch CF | 2.6 | 3.7 | 10.5 | % |
| Percent of Peak demand | 82 | 64 | 68 | % |
| Percent of Average demand | 108 | 104 | 110 | % |
| **Renewable Generation Use Parameters** | | | | |
| Renewable Gen | 258 | 518 | 277 | TWh |
| Curtailed Renew. | 35 | 126 | 75 | TWh |
| Percent Curtailed | 13 | 24 | 27 | % |

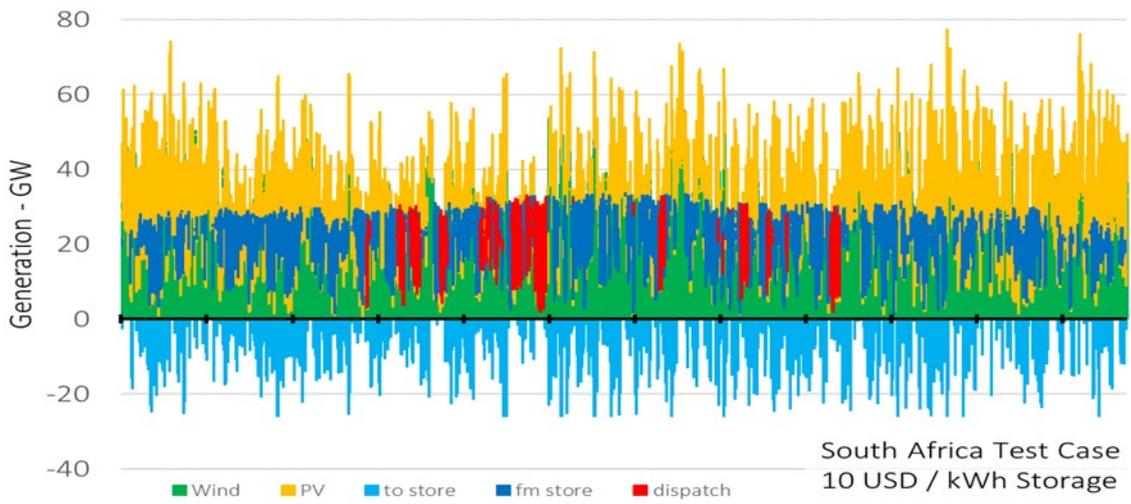

*Figure 7 - South Africa Test Case with 10 USD/ kWh storage*

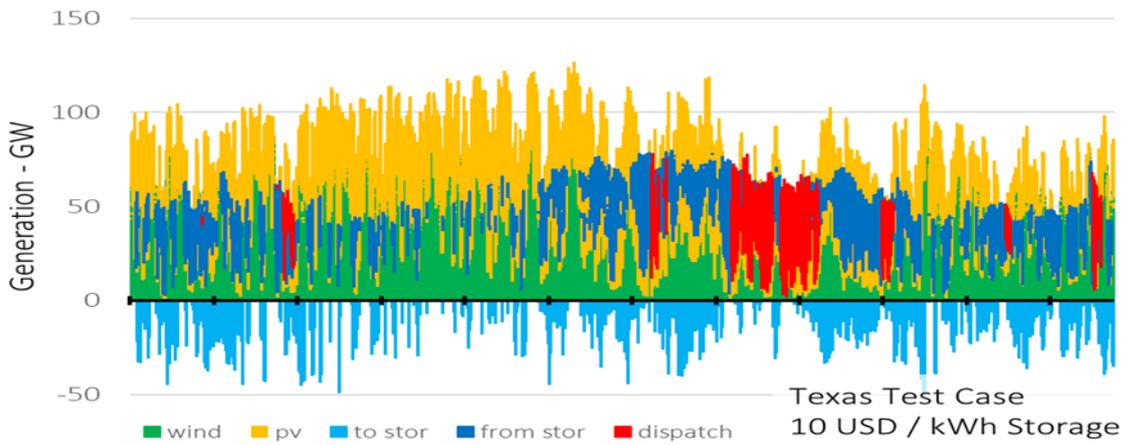

*Figure 8 - Texas Test Case with 10 USD/kWh Storage*

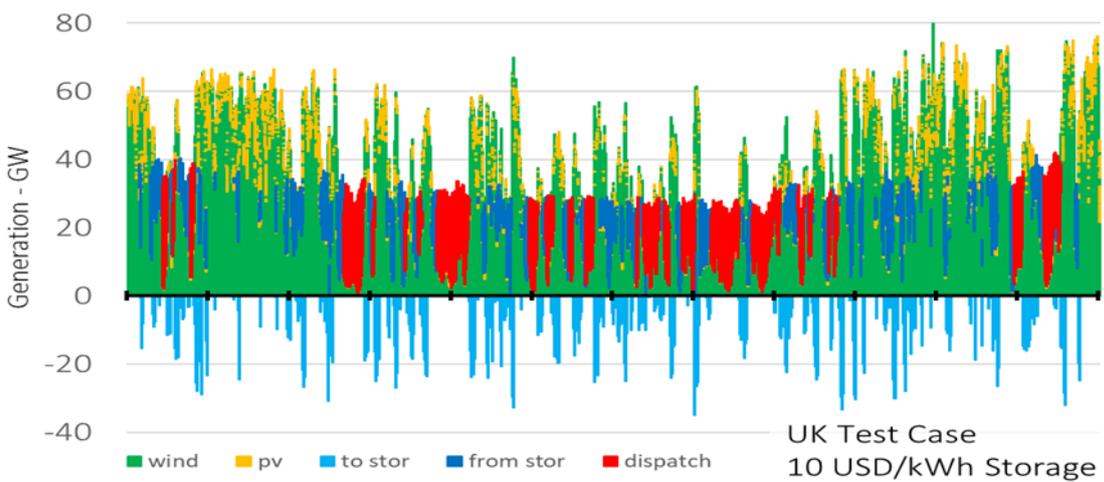

*Figure 9 - UK Case with 10 USD/kWh Storage*

**Low fuel cost**

While 20 USD/GJ fuel cost is appropriate for South Africa, it is likely to be overly high for the estimated fuel cost for Texas, where there is direct access for large volumes of natural gas and the existence of the gas distribution network to provide it to any power plant location. In the UK, as local natural gas production from the North Sea depletes, the market is increasingly supplied by LNG importation and the costs will approach those of the South African market. Using 10 USD/ GJ rather than 20 would change the cost of the use of firm-dispatchable power in these markets. The comparison of the Texas case with 20 USD / GJ to 10 USD/GJ indicates that the low fuel cost leads to a 5% increase in the installed firm-dispatchable generation but double the use for the lowest cost scenario.

**Cost Implications for Firm-dispatchable Power**

According to the latest estimates from the US EIA, generation costs for onshore wind is estimated to currently have a levelized cost of energy (LCOE) of 40 USD/per MWh and solar PV a LCOE of 36 USD per MWh, making them the lowest cost generation sources. This compares to an LCOE for new coal plants at 82 USD per MWh and for nuclear (conventional or SMR) at over 88 USD/ MWh (US EIA, 2023a). It is often argued that the requirement to add all the required firm-dispatchable generation to back up the system based on solar and wind generation makes the system unacceptably expensive. As this generation is only used less than 10% of the time, it becomes expensive power. While it is expensive to add the required firm-dispatchable generation, it should not add significantly to the cost of a wind and solar based system.

There are two costs that must be considered, the first is the cost of installing the required generation capacity and the second is the cost of providing the electricity from the installed capacity. With the assumed capital cost of 800 USD/kW for firm-dispatchable generation and an assumed thirty-year life, the amortization cost with an 8% interest rate for this would be 71 USD/a/kW. This cost for the amortization of the installation of the required firm-dispatchable generation in each of the three systems adds slightly less than 10 USD per MWh to the overall generation.

While the modelling was conducted with NREL forecast 2040 costs, even using the US EIA estimated capital costs for 2023, wind and solar have lower capital costs compared to any alternative technology (US EIA, 2023a). In each of the three scenarios modelled, the required firm-dispatchable generation is approximately 300 MW per GW of wind and solar generation in the system. Adding the capital cost for required backup with OCGT firm-dispatchable power does not change this comparison. This is shown in Figure 11. As can be seen in Figure 11, with current costs, wind and solar are the lowest capital cost technologies.

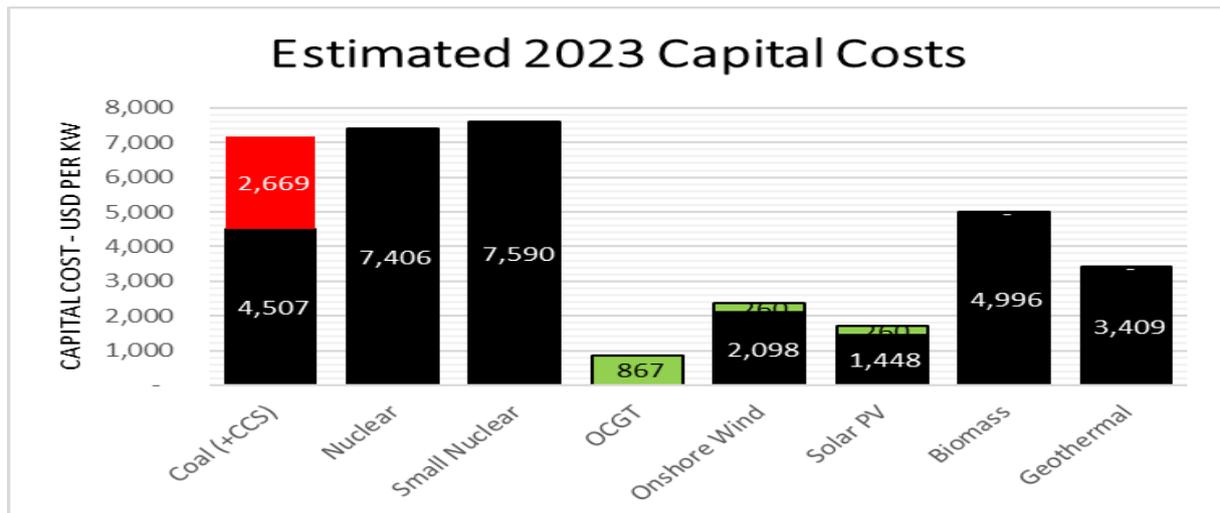

*Figure 10 - US EIA Estimated Capital Cost by Technology - data:* (US EIA, 2023a)

Usage Cost

For energy production, there is the fixed operating cost for the installed generation capacity plus the cost of fuel and some small amount of associated variable operating costs. The US EIA estimates the costs for generation by technology each year and the annual fixed costs for various types of dispatchable generation sources range from 7.88 USD / kW to 39.57 USD / kW, with OCGT on the low end of this range. (US EIA, 2023a). Using an estimated costs of 8 USD/kW to 40 USD/kW would add an additional 1 to 5 USD per MWh to the overall supply of electricity.

The variable cost was encompassed within the estimate of 20 USD / GJ which would equate to a cost of generation of approximately 200 USD /MWh on top of the capital and fixed operating costs. This compares to the total cost of 40 USD/ MWh for wind and the 36 USD/ MWh for solar PV. This high cost of generation indicates that their use should be minimized. Engines and turbines can be potentially fuelled with diesel, LPG, natural gas, biogas as well as hydrogen and its derivatives such as green ammonia (Siemens, 2020). Fuel cells are slightly more limiting in their fuel choices but can still use most of these (US Department of Energy, 2023). Most of the fuels should fall within this cost range and would be selected by the power plant based on the specific delivered cost at the power plant and other considerations. However, while the total fuel used in a year at a plant might not be large, its consumption during the usage will be quite significant and the availability of the fuel in the needed quantities for the usage period is a major consideration and one that can restrain the fuel choice (Clark et al., 2022; Clark, McGregor & Van Niekerk, 2022).

This cost for firm-dispatchable power could be covered by two market mechanisms and both are used around the world. These are capacity payment plus usage payment and an energy supply basis (Timera Energy, 2020). The capacity payment basis allows both the supplier and the user to go into the contractual arrangement with a clear understanding of their income stream where in the energy market the two parties must make their own assumption of how much energy their facility might provide and price the energy on this basis. Most utilities prefer to use the capacity payment system. Texas is one of the few systems that use an energy only payment system, but the conversion into a capacity plus system has often been discussed (US EIA, 2022; Wood Mackenzie, 2023)

With the assumptions made in this analysis, the capacity payment to the firm-dispatchable generator should be approximately 111 USD per year per installed kW. For all the cases studied, this would require a capacity payment of approximately 14 USD / MWh on the total supply. This would be plus cost of fuel when used, in the range of 200 USD/MWh.

**Summary of Results**

Scenarios have been tested for three markets with an assumed nominal 100 % supply from wind and solar resources. In all cases, there is a requirement for installed backup generation meeting approximately 110 % of the average annual demand. The usage of this installed firm-dispatchable generation capacity varies significantly depending on the costs of wind, solar and storage. The use of the firm-dispatchable generation also varies with the cost of fuel. However, the amount of firm-dispatchable generation capacity that must be installed is consistent within all scenarios. This was tested both compared to the varied supply and demand profiles in the three locations and with the variations of fuel and storage costs.

While it is appropriate to use the assumption of systems based effectively completely on wind and solar for comparison purposes, this is not reality. Each of these regions has legacy firm generation that will likely to be economical for some years to come and the renewable generation sources will only partially displace the firm generation for as long as they are economical to use. For this analysis, it was assumed that 10 GW would remain in South Africa and the UK and 20 GW in Texas. These scenarios were covered for the three regions in the analysis shown in supplement 2. As can be seen from this analysis, the remaining firm generation will effectively directly reduce the volume of total firm-dispatchable generation but not affect its requirement proportionally to the variable generation. Retaining 10 GW with an assumed 60% EAF would displace 6 GW of the requirement for firm-dispatchable generation.

In the three models, the required installed firm-dispatchable generation was found to meet the majority but not all the peak demand (68% to 84%). This premise would need to be checked with multi-year modelling as it is possible that in some years the lack of wind and solar might correlate closer with the peak demand. However, it would also be possible to balance this peak with demand response rather than additional firm-dispatchable generation.

The cost of the required firm-dispatchable generation will add to the overall cost of the wind and solar based system. In all three of the test markets, the installation of this firm-dispatchable generation capacity would add approximately 14 USD / MWh to the cost of the total generation. In addition, depending on fuel costs, the energy from this would cost approximately 200 USD /MWh. This high cost of energy would lead towards minimal use of this generation capacity.

**Conclusion**

As noted in the introduction, reports have indicated that a generation system based on wind and solar generation is the lowest cost way of building a sustainable generation system in South Africa and internationally. However, the studies have noted the variable nature of these resources and concluded that "peaking power" must be added to balance supply and demand. As this research has shown, the need is not as much for peaking power but for firm-dispatchable power. This is generating capacity that is always available to completely replace the wind and solar sources and that can be turned on or off, or can adjust its power output according to market need. Energy storage will reduce or replace the need for peaking power, but even large amounts of energy storage will not change the requirement for the installation of firm-dispatchable generation. The amount of energy that this required backup generation will provide will be dictated by the economics of supplying power from the various sources. However, the requirement for installing enough firm-dispatchable energy to completely replace the wind and solar generation is essential to keeping the system in balance.

**Further Research**

As noted in this analysis, the comparison was with three isolated network systems and not with systems integrated into extensive networks that might provide balancing to mitigate the firm-dispatchable energy requirement to some extent. While complex, this is analysis that should be done. As noted above, the use of this firm-dispatchable generation is minimized to keep the cost low. Current fuels will have some greenhouse gas emissions but are a secondary target. Eventually, economics and policies will dictate the changeout of the fuel used to non-greenhouse gas emitting fuels to achieve net zero targets. This could be in the form of biofuels, but it can also be sourced from green hydrogen produced by excess renewable generation. The economic and technical analysis of this option will be pursued.

**Conflict of Interest**

The authors have no known conflicts of interest.

**Supplement 1**

**Systems with 100 % PV generation and battery storage**

It has been argued that, with the declining cost of solar PV and batteries, it should be possible to completely supply the electricity needs of the country or region using solar generation with battery storage (Dorr & Seba, 2020). The amount of solar power that must be installed is not a matter of cost, only a matter of the capacity factor and timing of the solar generation. For the three test regions, South Africa, Texas and the UK, the requirements were tested in the model. As can be seen in Table S1-1, the amount of required installed solar is quite significant, at 570% of peak demand for South Africa, 908% of peak demand in Texas and 2140% of peak demand in the UK. In addition, installed battery capacity must be installed to more than completely meet the peak demand for several days – 24 hours for South Africa and up to 44 hours in Texas. The overbuilding of the battery storage is required to provide for the round-trip losses from energy put into the storage and taken out for reuse. As can be seen from the table, the solar generation will also have a significant amount of curtailed generation, from 47% in South Africa to approximately 70% in Texas and the UK. Some of this excess energy might be utilized for applications that can handle the variable supply.

The cost for this large installation of solar plus batteries is undefined in the modelling, but even with optimistic assumptions, it is likely to be more expensive than a system that allows some of the backup generation to be met from firm dispatchable power with much less solar and battery overbuilding. As shown in the base case analysis, solar plus wind resources would need to be 258 % of the peak demand in South Africa, 242 % in Texas and 231 % in the UK in the minimum cost case with firm dispatchable generation. The additional 312 % for South Africa, 666 % for Texas and 1909 % of peak demand for solar generation plus the additional batteries would be the required installation only to replace the firm dispatchable generation.

**Rigidity Test**

With the amount of solar PV installed as noted above, it is possible to meet the nominal needs for the generation system. For each modelled scenario, an actual weather year was utilized with the relevant generation and demand. However, there are many potential factors that change the balance in the system, whether it is weather effects that lead to lower generation, systems that do not perform as designed or demand that might increase due to specific undefined events. To see how these three systems respond to unknown events, the demand profiles were increased in the model until the installed solar and batteries do not meet the demand. As can be seen in Table S1-2, only an extremely small change is required in the demand to completely change the system requirements. For South Africa, this occurred when the demand was increased by 3% and for Texas and the UK, this occurred when the demand increased by 1% or less.

*Table S1-1 - Cases with Complete PV and Batteries*

|  | South Africa-Eskom | Texas - ERCOT | UK - NG | units |
|---|---|---|---|---|
| **Scenario Parameters based on 2022** | | | | |
| Annual Demand | 231 | 431 | 230 | TWh |
| Peak Rate | 34 | 80 | 43 | GW |
| Average Rate | 26 | 49 | 26 | GW |
| **Parameters for complete Supply from Solar PV plus batteries** | | | | |
| Installed PV | 195 | 726 | 920 | GW |
| PV Energy | 430 | 1492 | 4 | TWh |
| PV CF | 26 | 23 | 9 | % |
| PV Percent of Peak Capacity | **570** | **908** | **2140** | Percent of Peak Gen. Capacity |
| Battery Capacity | 51 | 76 | 68 | GW |
| Battery Hours | 24 | 44 | 30 | Hours |
| Battery Energy | 1224 | 3344 | 2040 | GWh |
| **Solar Generation Use Parameters** | | | | |
| Renewable Gen | 430 | 1492 | 765 | TWh |
| Curtailed Renew. | 209 | 1061 | 526 | TWh |
| Percent Curtailed | 47 | 71 | 69 | % |

The failure mode in these cases is as significant as the small value of the increase before the system fails to meet the needs. In all cases, these small changes require that firm dispatchable power is installed to meet the total demand. While it might only be required for a very minimal time, without it the grid would collapse. Building a system with this rigidity would not be practical. The failure modes for these scenarios in the three systems are demonstrated in Figures 1, 2 and 3. The required firm generation needed to balance the system is highlighted within the red ellipses. These figures seem quite insignificant compared to the vast amount of solar generation and the amount of storage energy that is used in these scenarios, but this system failure case is a major concern.

While it would be possible to, at some cost, build a system to encompass any expected increases above the nominal expected load, the rigidity in the system remains. The failure mode has the same major negative consequence should the revised limit be breached even by a minimal amount.

*Table S1-2 - Test Cases for Demand Changes to Scenarios of 100% PV and Batteries*

|  | South Africa-Eskom | Texas - ERCOT | UK – NG | units |
|---|---|---|---|---|
| **Test for Excess Demand in 100% Solar PV Scenario** | | | | |
| Annual Demand | 231 | 431 | 230 | TWh |
| Test Demand | 239 | 434 | 232 | TWh |
| Percent of Normal | 103 | 101 | 101 | % |
| **Firm Dispatchable Energy Needed to Balance for Excess Case** | | | | |
| Installed Dispatch | 24 | 56 | 27 | GW |
| Dispatch Energy | 298 | 242 | 13 | GWh |
| Percent of Average demand | 87 | 113 | 104 | % |

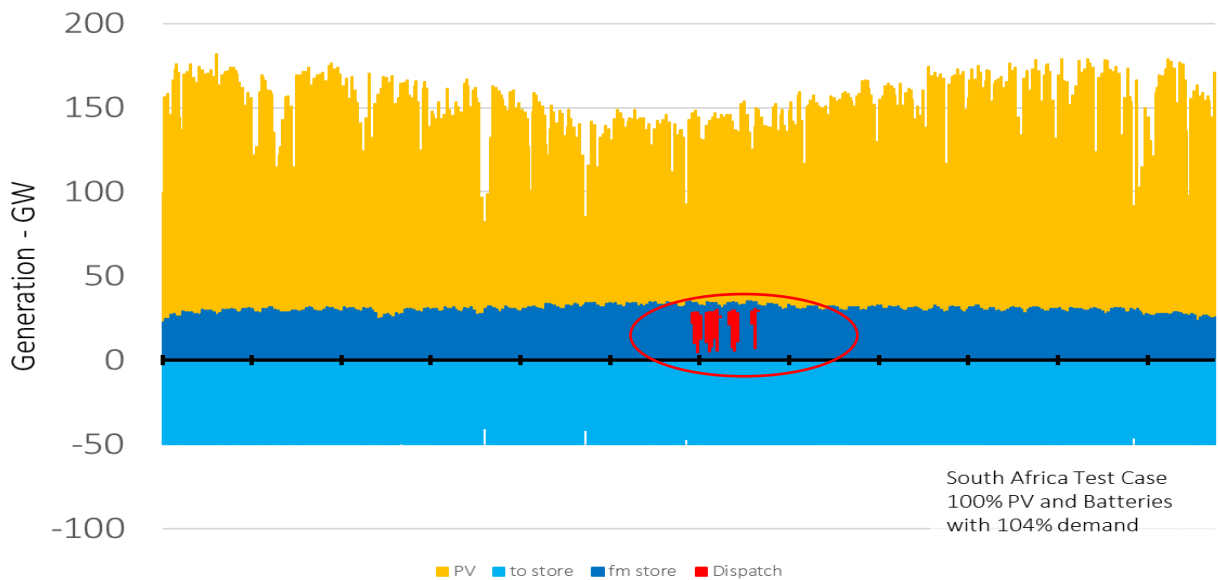

*Figure S1-1 - South Africa Test with 100% PV and Batteries*

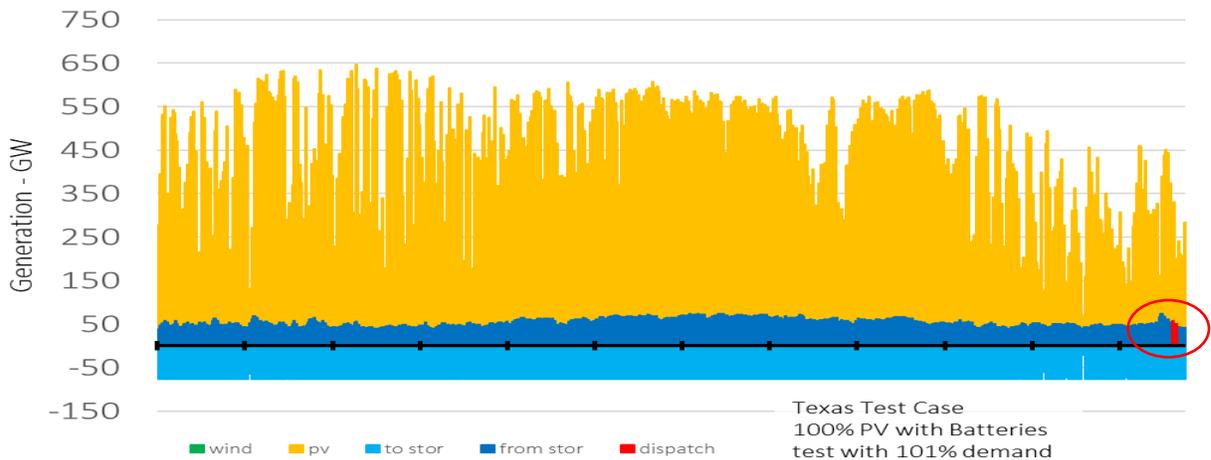

*Figure S1-2 - Texas Test with 100% PV and Batteries*

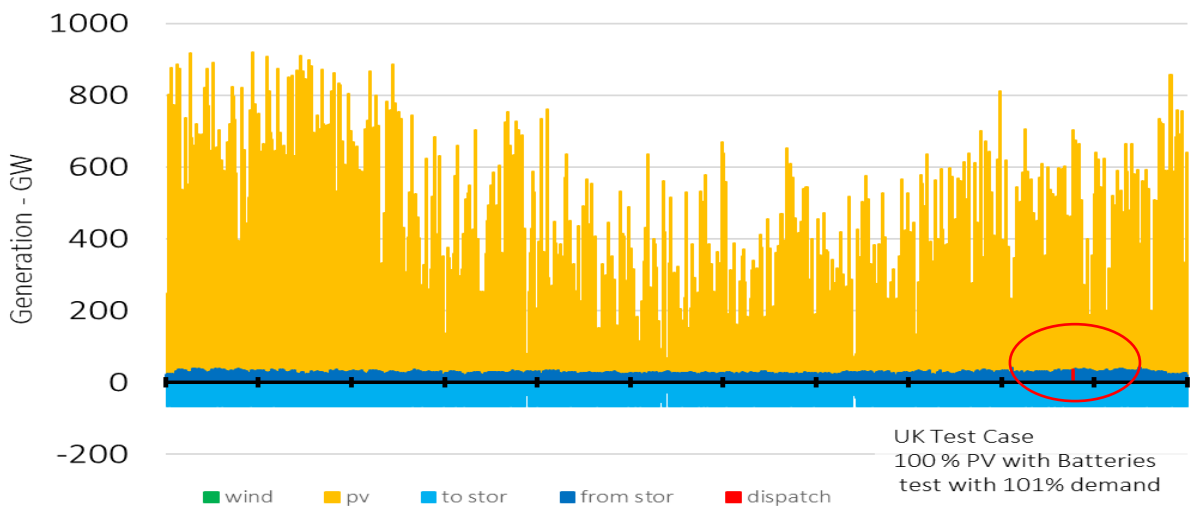

*Figure S1-3 - UK Test with 100% PV and Batteries*

## Conclusion

From the results of these simulations, it must be concluded that a system without the provision for firm dispatchable generation, while theoretically possible, is expensive and overly rigid. This is not a scenario that should be pursued in any system.

## Reference

Dorr, A. & Seba, T. 2020. Rethinking Energy 2020-2030 100% Solar, Wind, and Batteries is Just the Beginning. *Circular economy*. 4(October):1–12. Available: https://www.systemiq.earth/wp-content/uploads/2020/12/The-Paris-Effect_SYSTEMIQ_Full-Report_December-2020.pdf%0Ahttp://teknologiateollisuus.fi/en/focus/environmental-issues/circular-economy%0Ahttps://www.vtt.fi/sites/handprint/PublishingImages/Carbon_Hand.

**Supplement 2**

**Systems with Residual Base Load Generation**

The base cases for the three scenarios assumed that 100% of the generation was from a system anchored on wind and solar generation, with no base load generation. In all realistic scenarios, there is a probability that some of the base load generation plants will not have reached their economic limits and it would be a lower cost to use these facilities up to those limits rather than replace them. To understand the impact of having some base load generation, new cases were developed with some remaining base load generation.

One of the major questions about the remaining base load generation is how it will be utilised. As wind and solar PV have no fuel cost, their incremental usage has minimal cost. On the other hand, base load generation from fossil fuel thermal plants – particularly coal fuelled plants, will have a significant cost for fuel. This fuel cost will make it more expensive to use this generation compared to wind and solar. Thermal plants are slower to ramp up and ramp down which makes them challenging to use in balancing variable supply and demand imbalances (Nichols, 2016). In addition, the added stress from varying loads increases the wear on mechanical systems, reducing their life expectancy (Bergh & Delarue, 2015). These considerations will make the decisions on when to maximize the use of these generation units a complicated decision.

For the sake of this comparison, the assumption is that base load generation will be the first utilised and therefore will be used at its full potential for the entire year and any curtailing will be done from the wind and solar generation facilities. As intent of the analysis is only to find out the impacts on the required firm-dispatchable generation, this assumption should not materially affect the results. The test if for the requirement for additional generation for the firm-dispatchable generation and what happens with excess potential generation should not impact this requirement.

For South Africa and for the UK, the case studied assumes that 10 GW of existing base load generation remains in the system. For Texas, as it is almost twice the size of the other two regions, the assumption was that 20 GW of base load remains. No changes to the demand profiles or the generation profiles from the wind and solar PV were considered in this analysis. The results from the three regions are shown in Table S2-1 and Figures S2-1 through S2-3.

*Table S2-1 - Scenarios with Residual Base Load Generation*

|  | South Africa-Eskom | Texas - ERCOT | UK - NG | units |
|---|---|---|---|---|
| **Current Parameters based on 2022** | | | | |
| Annual Demand | 231 | 431 | 230 | TWh |
| Peak Rate | 34 | 80 | 43 | GW |
| Average Rate | 26 | 49 | 26 | GW |
| **Lowest Cost Case Parameters with Residual Baseload Generation** | | | | |
| Base load Gen. | 10 | 20 | 10 | GW |
| Base EAF | 70 | 70 | 70 | % |
| Base Energy | 61 | 122 | 61 | TWh |
| Installed Wind | 62 | 107 | 98 | GW |
| Wind Energy | 197 | 323 | 131 | TWh |
| Wind CF | 36 | 34 | 31 | % |
| Wind Percent of Peak Capacity minus net base | 229 | 163 | 272 | Percent of Peak Gen. Capacity |
| Installed PV | 17 | 57 | 5 | GW |
| PV Energy | 38 | 117 | 2 | TWh |
| PV CF | 26 | 23 | 9 | % |
| PV Percent of Peak Capacity | 63 | 86 | 14 | Percent of Peak Gen. Capacity |
| Battery Capacity | 6 | 12 | 0 | GW |
| Battery Hours | 4 | 3 | 0 | Hours |
| Installed Dispatch | 22 | 46 | 23 | GW |
| Dispatch Energy | 10 | 23 | 24 | TWh |
| **Dispatchable Generation Parameters for Lowest Cost Case with Base Load** | | | | |
| Dispatch CF | 5.2 | 5.6 | 11.8 | % |
| Percent of net Peak demand | 84 | 74 | 64 | % of net Peak |
| Comparison from Base Case | 84 | 73 | 60 | % of total Peak |

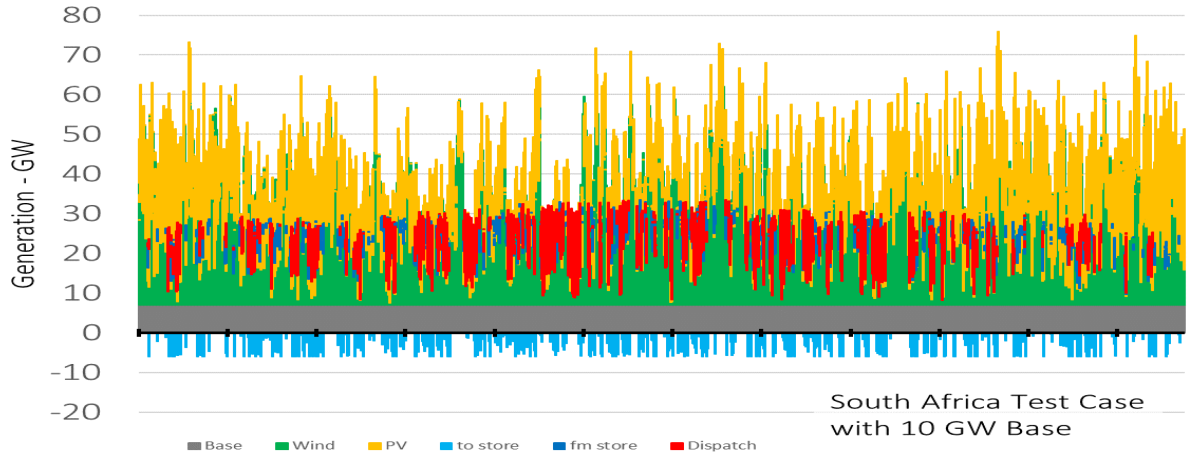

*Figure S2-1 - South Africa Test Case with Base Load*

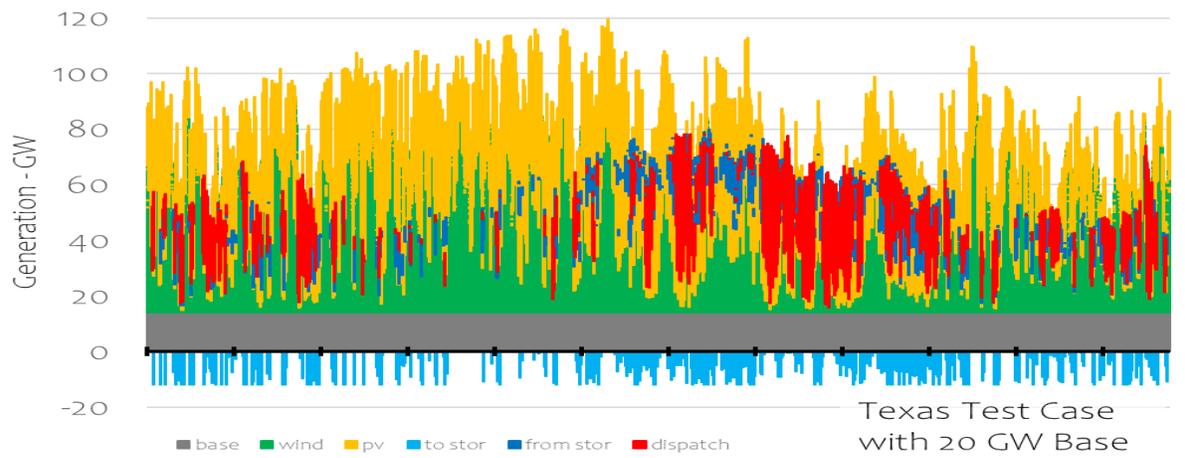

*Figure S2-2 - Texas Test Case with Base Load*

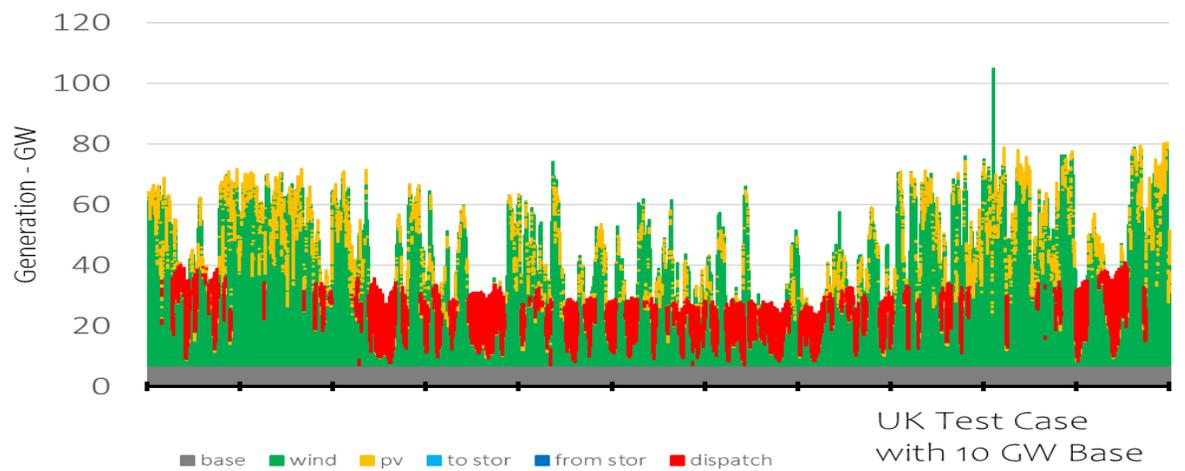

*Figure S2-3 - UK Test Case with Base Load*

**Results**

As can be seen from the results from the three cases, having a portion of residual base load generation will impact the requirement for firm-dispatchable generation in direct proportion to the amount of the total load that is covered by the base load generation. Firm-dispatchable generation must be available to completely replace the variable generation portion of the supply. In all cases, the requirement with respect to the proportion of the net peak demand was unchanged as demonstrated in the final two lines in Table S2-1.

**Conclusion**

The results confirm that some form of firm generation must be available to provide for the entire demand profile. Firm-dispatchable is required to completely support the portion of the demand that is being met by the variable supply. This would imply that firm-dispatchable generation must be built into the system at the same rate that base load generation is removed through the decommissioning of the existing thermal plants.